\documentclass[12pt,twoside]{article}
\usepackage[dvips]{graphicx}
\usepackage{color}

\newcommand{\FD}{\ensuremath{F_{2}^{D(3)}}}
\newcommand{\dcs}{\ensuremath{\sigma}}
\newcommand{\aem}{\ensuremath{\alpha_{em}}}

\newcommand{\be}{\begin{equation}}
\newcommand{\ee}{\end{equation}}

\begin{document}
\begin{flushright}
%MC-TH-**-**\\
\end{flushright}
{\large  Gluon Saturation in the Colour Dipole Model?} \\

\noindent J.~R.~Forshaw and G.~Shaw$^{a}$

\vspace*{0.5cm}

\noindent $^{a}$Particle  Physics Group, School of Physics and Astronomy,\\
The University of Manchester, M13 9PL, UK
\vspace*{0.5cm}

\begin{abstract}
We use data on the deep inelastic structure function $F_2$ in order to constrain
the cross-section for scattering a colour dipole off a proton. The data seem to
prefer parameterisations which include saturation effects. That is they indicate
that the strong rise with energy of the dipole cross-section, which holds for
small dipoles, pertains only for $r < r_s(x)$ where $r_s(x)$ decreases monotonically as
$x$ decreases. Subsequent predicitions for the diffractive structure function
$F_2^{D(3)}$ also hint at saturation, although the data are not really sufficiently
accurate.   
\end{abstract}
\section{Introduction}

The possible observation of gluon saturation effects  in the HERA
data at very low $x$-values  has been discussed extensively in the context of
the colour dipole model. In  particular,  the ``saturation model'' of 
Golec-Biernat and W\"usthoff~\cite{GW99a,GW99b} was shown some years ago to
give an elegant and accurate account of deep inelastic scattering (DIS) at small $x$, 
and a rather good description  of the  diffractive deep inelastic scattering (DDIS)
data~\cite{H197,ZEUS99}. More recently, a new saturation model,
the Colour Glass Condensate (CGC) model of  Iancu, Itakura and Munier
\cite{IIM04}, has been formulated. This model, can be  thought of as a more sophisticated version
of the Golec-Biernat--W\"usthoff model. Gluon saturation
effects are now incororated via an approximate solution of the
Balitsky-Kovchegov equation \cite{BK96}, which applies in the perturbative
region when the gluon densities become large and non-linear effects become important.
The resulting dipole cross-section,
obtained by fitting the free parameters of the model to the DIS data
at low $x$, is very similar to that obtained in the original saturation
model\footnote{For an explicit comparison of the two dipole cross-sections, see \cite{FSS04a}.}
and Forshaw, Sandapen and Shaw \cite{FSS04b} have shown that it leads to similar, successful,
predictions for the DDIS data.

It is clear from this and other work \cite{satgen} that the predictions of saturation models
are compatible with a wide range of data. 
However, about the same time as the saturation model was formulated, Forshaw,
Kerley and Shaw (FKS) proposed a  two-component Regge dipole model \cite{FKS99} which
explicitly excludes gluon saturation effects. As in the saturation model, the
parameters of the model were tuned to fit the DIS data at low-$x$, and were
subsequently shown \cite{FKS2000} to give a good account of the DDIS data
without further adjustment. Subsequently, the predictions of all three models
for the DDIS data have been compared to each other \cite{FSS04b}, showing that
they were
all in good agreement with the data. However, at energies just above the
measured region, the predictions of the saturation models and the unsaturated
two component model rapidly diverge. A similar pattern emerges in deeply
virtual Compton scattering \cite{FSS04c} and vector meson production \cite{FSS04a},
although in the latter case, there are large ambiguities associated with the
wavefunctions of the vector mesons.

From this it is clear that the data analysed to date are insufficient to establish the
existence of gluon saturation. However, since the saturation model and the
two-component Regge model were originally formulated, much more precise data have become
available on both the DIS structure function $F_2$ \cite{newf2data} and the DDIS 
structure function $F_{2}^{D3}$ \cite{H103}. In this
paper we will analyse these new data to see if they can throw more light on the
question of gluon saturation.

\section{Colour Dipoles and Saturation}

\begin{figure}[htb]
\begin{center}
\includegraphics[width=8cm]{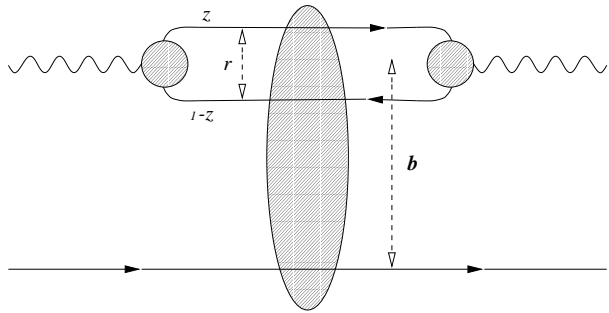}
\caption{The colour dipole model for $\gamma^* p \to \gamma^* p$.}
\label{f2dipole}
\end{center}
\end{figure}

In the colour dipole model ~\cite{NZ91,Mueller94},  the
forward amplitude for virtual Compton scattering is assumed to be dominated 
by the mechanism illustrated in Figure \ref{f2dipole} in which the photon fluctuates into a
$q \bar{q}$ pair of fixed transverse separation $r$ and the quark carries a 
fraction $z$ of of the incoming photon light-cone energy. 
Using the optical theorem, this leads to 
\begin{equation}
\sigma^{L,T}_{\gamma^{*}p} = \int dz \;  d^2 r \ 
|\psi_{L,T}(z,r)|^{2} \dcs(s^*,r)  
\label{dipoledis}
\end{equation} 
for the total virtual photon-proton cross-section, where $
\psi_{L,T}(z,r)$ are  the  appropriate spin-averaged light-cone 
wavefunctions of the photon and 
$\dcs(s^*,r)$ is the dipole cross-section. The dipole cross-section is usually assumed to be
independent of $z$ (see below), and is parameterized in terms
of an energy variable $s^*$ which depends on the model. 

%models  
%the squared centre of mass energy $s$, while others
%choose the Bjorken scaling variable $x$. This latter choice precludes the
%cosideration of the energy dependence of real photon processes, since $x=0$
%independent  of energy at $Q^2 = 0$. 
%For this reason, a third alternative is a
%modified  scaling variable which reduces to $x$ at large $Q^2$  but  $ ~ s^{-1}$ as $Q^2
%\rightarrow 0$, for example 
%\be
%x_f = \frac{Q^2}{Q^2+W^2} \left( 1 + \frac{4m^2}{Q^2} \right) 
%\ee
%where $m_f$ is the relevant quark mass.

In the dipole model, not only the DIS total cross-section, but also
the forward amplitudes for DDIS and DVCS, are determined by the
same ingredients: the light-cone
wavefunctions of the photon and the dipole cross-section. We shall briefly
discuss each in turn. 

For small $r$, the light-cone photon wavefunctions  are
given by the tree level QED expressions~\cite{NZ91,DGKP97}:
\begin{eqnarray}
  \label{eq:psi^2}
  |\psi_{L}(z,r)|^{2} & =  & \frac{6}{\pi^{2}}\aem\sum_f e_{f}^
{2}Q^{2}z^{2}(1-z)^{2} K_{0}^{2}(\epsilon r) \\
  |\psi_{T}(z,r)|^{2} & = & \frac{3}{2 \pi^{2}}\aem\sum_{f} e_{f}^
{2} \left\{[z^{2} + (1-z)^{2}] \epsilon^{2} K_{1}^{2}(\epsilon r) + m_{f}^{2} 
K_{0}^{2}(\epsilon r) \right\} 
\end{eqnarray}
where 
\(
 \epsilon^{2} = z(1-z)Q^{2} + m_{f}^{2}\; ,
\) 
$K_{0}$ and $K_{1}$ are modified Bessel functions and the sum is over 
quark flavours $f$ with quark masses $ m_f$. These expressions decay
exponentially at large $r$, with  typical $r$-values of order $Q^{-1}$ at large
$Q^2$ and of order $ m_f^{-1}$ at $Q^2 = 0$.  
However for  large dipoles\footnote{The contributions from very large dipoles $r
\gg 1$ fm are heavily suppressed at all $Q^2$ because of the exponential decay of the wavefunction
at large $r$.} $r \ge 1$fm, which are  important at low $Q^2 \le 4
m_f^2$, a perturbative treatment is not appropriate, and hadronic states of
this transverse size  are more sensibly modelled as pairs of constituent quarks,
or vector mesons. For this reason, FKS \cite{FKS99,FKS2000} used constituent quark
masses, together with  an
enhancement factor in the region $r \approx 1 $ fm, as suggested by Generalised
Vector Dominance (GVD) ideas \cite{FGS98,GVD1,GVD2}. In contrast, the saturation models retain the
perturbative wavefunction, but use a much smaller quark mass, which again
enhances the wavefunction at large $r$. We have investigated both these
approaches and find that the difference between them 
is only important when confronting the photoproduction data (mainly that from fixed-target
experiments \cite{Caldwell78}). Since we are primarily interested in
determining the need for saturation effects we will not need to consider these data here, 
and will therefore use the pertubative QED wavefunction (\ref{eq:psi^2}) throughout,
treating the quark mass as a parameter which can be used
to adjust the wavefunction at large $r$-values. At high $Q^2$ where small
dipoles dominate, the wavefunctions are insensitive to the quark mass.

Turning to the dipole cross-section, it is useful to distinguish three regions:
small $r$, where pertubative ideas are relevant; large $r$, where 
typical ``hadronic'' behaviour is expected; and intermediate dipole sizes.
The dipole formula (\ref{dipoledis}) can be derived in the  
leading $\log(1/x)$ approximation of perturbative QCD \cite{FrankRS}.  In this
approximation, the wavefunction is given by the perturbative expression 
 (\ref{eq:psi^2}) and the dipole cross-section is given by
\be
\sigma (s^*,r) = \frac{4\pi^2}{3}\alpha_{s}\int \frac{dk^2}{k^4} f(x,k^2)(1-J_0(kr))
\sim \frac{\pi^2 r^2}{3} \alpha_s xg(x,A^2/r^2),
\label{pertdcs}
\ee
where $f(x,k^2)$ is the unintegrated gluon density and the approximate equality
holds for sufficiently small $r$ (with $A \sim 3$).
As $x$ gets smaller, the gluon distribution grows rapidly, 
and the dipole cross-section in this region is sometimes approximated by the 
``hard  pomeron'' behaviour
\be
\sigma (s^*,r) \approx a r^2 x^{-\lambda_H}
\label{hardpom}
\ee
with $\lambda_H \approx 0.4$.
At large $r \ge 1$ fm, which is important as $Q^2 \to 0$, the models we discuss
all build in ``soft pomeron'' behaviour 
\be
\sigma (s^*,r) \approx \sigma_0 s^{\lambda_S} \; , 
\label{softpom}
\ee  
with  $ \lambda_S$ small or  zero.
 The dipole cross-section is then interpolated
between the two in the intermediate $r$ region, in a way that depends on the
specific model. 
For example, in  the original Regge dipole model model of Forshaw, Kerley and 
Shaw \cite{FKS99,FKS2000},  the dipole
cross-section was extracted from DIS and real photoabsorption data assuming a form with two 
terms, each with a  Regge inspired $s$ dependence:
\begin{equation}
\label{gammatot}
  \sigma(s,r)  =   a_{soft}(r) s^{\lambda_{S}} + a_{hard}(r) s^{\lambda_{H}} 
\; 
\label{fkssigma}
\end{equation} 
where the values $\lambda_S \approx 0.06$, $\lambda_H \approx 0.4$
resulting from the fit are characteristic of soft and hard diffraction
respectively. The functions $a_{soft}(r)$, $a_{hard}(r)$ were chosen so
that for  small dipoles the hard term dominates yielding a 
behaviour $\sigma \rightarrow r^2 (r^2 s)^{\lambda_H}$ as $r \rightarrow 0$.
This in accordance with the colour transparency ideas embodied in (\ref{pertdcs}), 
since the dimensionless
variable $r^2 s$ is closely related to $x$ at large $Q^2$, 
where the typical dipole sizes $ r \propto Q^{-1}$. For large dipoles
 $r \ge 1$ fm the soft 
term dominates with a hadronlike behaviour $\sigma \approx \sigma_0 
(r^2 s)^{\lambda_S}$. 

In any model which incorporates the ``hard pomeron''
behaviour (\ref{hardpom}), the dipole cross-section  increases rapidly with
decreasing $x$ for the small dipole sizes which dominate at large $Q^2$.
In nature, however, one might
expect this sharp rise  to be softened  by unitarity effects, an effect which
we shall refer to as ``gluon saturation'' since, if it occurs in
the region where (\ref{pertdcs}) is valid, it can be interpreted as a dampening
of the rapid increase of the gluon density with decreasing $x$ implied by 
(\ref{hardpom}). Such effects are not included in the FKS model
\cite{FKS99,FKS2000}, where for a small dipoles of a given $r$, the behaviour
(\ref{hardpom}) holds indefinitely as $x \rightarrow 0$.\footnote{This is not to deny
the inevitable role of nonlinear dynamics, merely to imply that it may not be needed in the
range of existing data.} 
Non-linear saturation dynamics is, however, explicitly incorporated into the CGC model,
in which the dipole cross-section is assumed to be of the form \cite{IIM04}
\begin{eqnarray}
\sigma(x,r) &=& 2 \pi R^2 {\cal N}_0 \left( \frac{r Q_s}{2} 
\right)^{2\left[\gamma_s + \frac{\ln(2/rQ_s)}{\kappa \lambda \ln(1/x)}\right]} 
\hspace*{1cm} \mathrm{for} \hspace*{1cm} rQ_s \le 2 \nonumber \\
&=& 2 \pi R^2 \{1 - \exp[-a \ln^2(brQ_s)]\} \hspace*{1cm} \mathrm{for} 
\hspace*{1cm} rQ_s > 2~,
\label{cgc-dipole}  
\end{eqnarray} 
where the saturation scale $Q_s \equiv (x_0/x)^{\lambda/2}$ GeV and
$x=Q^2/(Q^2+s)$ is the Bjorken scaling variable. 
The parameters are fixed by a combination of theoretical constraints and a fit
to DIS data\footnote{The authors of \cite{IIM04} carried out a series of fits for different
values of the parameter $N_0$, and found that the results were only
weakly dependent on this parameter. In this paper we choose the fit corresponding to
$N_0 = 0.7$.}. This dipole cross-section is characterised by a rapid increase
with decreasing $x$, not dissimilar to (\ref{hardpom}), at small $r$, changing
to a softer energy dependence as $r$ increases beyond 
$$
 r_s(x) \equiv 2/Q_s = \; 2 (x/x_0)^{\lambda/2} \; .
$$ 
Saturation arises because of the decrease of the ``saturation radius'', $r_s$, with
decreasing $x$. Let us consider a dipole of fixed transverse size $r$. 
If $r < r_s(x)$, the dipole cross-section 
 increases rapidly
as $x$ decreases. However, this rapid rise eventually switches to a softer
$x$ dependence when $x$ becomes so small that $r_s(x)$ itself decreases below
the the fixed dipole size $r$.

%\begin{figure}[htb]
%\begin{center}
% \includegraphics*[width=8cm]{}\includegraphics*[width=8cm]{}
%\caption{Left: the FKS dipole cross-section for  $W = 10,75$ and $300$ GeV$^2$. 
%Right: the CGC dipolev cross-section at **** 3 suitable x-values} 
%\label{sigmadipole}
%\end{center}
%\end{figure}

Finally we remark that the word ``saturation'' is also often used to describe a
somewhat different effect. In all models, the
approximate $r^2$ dependence which pertains in the perturbative region at 
small $r$ is flattened when $r$  increases to  values in the
non-perturbative region. This ``non-perturbative saturation'' is important in
describing the transition from high to low $Q^2$ values. However it is not
what we are concerned with here. 

\section{Analysis of the HERA data}

The parameters of the original Regge dipole model of FKS were determined using the DIS data 
available in 1999. However, since then more precise measurements of the deep inelastic
scattering data in the diffractive region have been made at HERA \cite{newf2data}, and
there exists new data on DDIS \cite{H103}.
In this section, we examine these newer data to see whether they can throw any new light
on the issue of gluon saturation. 

Our strategy is to confront the HERA data with a new Regge dipole model (which can be
thought of as an update of the original FKS model to accommodate the latest data) 
and also a variant of that model which includes saturation. 
We shall also always compare to the predictions of the CGC saturation model. 

Let us introduce our new and very simple Regge inspired dipole model.
We shall assume that 
\begin{eqnarray}
\sigma(x_m,r) &=& A_H r^2 x_m^{-\lambda_H}~~{\mathrm{for}}~~r < r_0~~{\mathrm{and}} \nonumber \\
 &=& A_S x_m^{-\lambda_S}~~{\mathrm{for}}~~r > r_1, \label{eq:FS04}
\end{eqnarray}
where 
\be
x_m = \frac{Q^2}{Q^2+W^2} \left( 1 + \frac{4m^2}{Q^2} \right).
\ee
For light quark dipoles, the quark mass $m$ is a parameter
in the fit, whilst for charm quark dipoles the mass is fixed at 1.4~GeV.

In the intermediate region  $r_0 \leq r \leq r_1$ we interpolate linearly between the
two forms of (\ref{eq:FS04}). 
Whether this is a Regge inspired model or a saturation model depends entirely upon the
way in which the boundary parameter $r_0$ is determined.

\subsection{Regge Fit to DIS data}

If the boundary parameter $r_0$ is kept constant then the parameterisation reduces
to a sum of two powers, as might be predicted in a two pomeron approach. 
It is plainly unsaturated, with the dipole cross-section obtained
at small $r$-values growing rapidly with increasing $s$ at fixed $Q^2$ (or
equivalently with decreasing $x$) without damping of any kind.

In all cases, we shall fit the electroproduction data in the kinematic 
range 
\be
0.045\, {\rm GeV}^2 < Q^2 < 45 \, {\rm GeV}^2  \hspace{0.5cm}   x \le 0.01 \; ,
\ee 
which is identical to that which was used to determine the parameters of the
CGC saturation model.

\begin{figure}[htb]
\begin{center}
\includegraphics*[width=7cm]{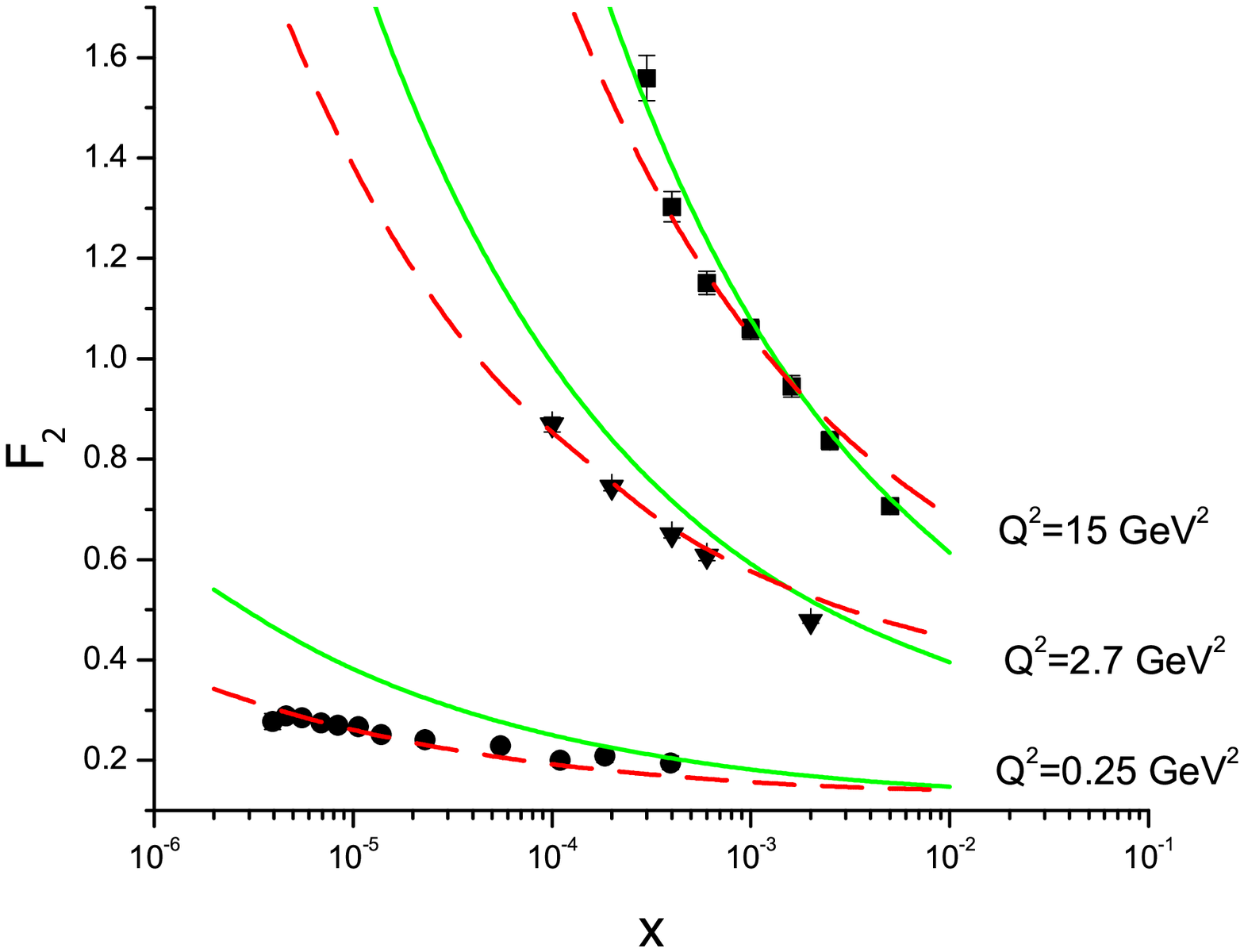}\includegraphics*[width=7cm]{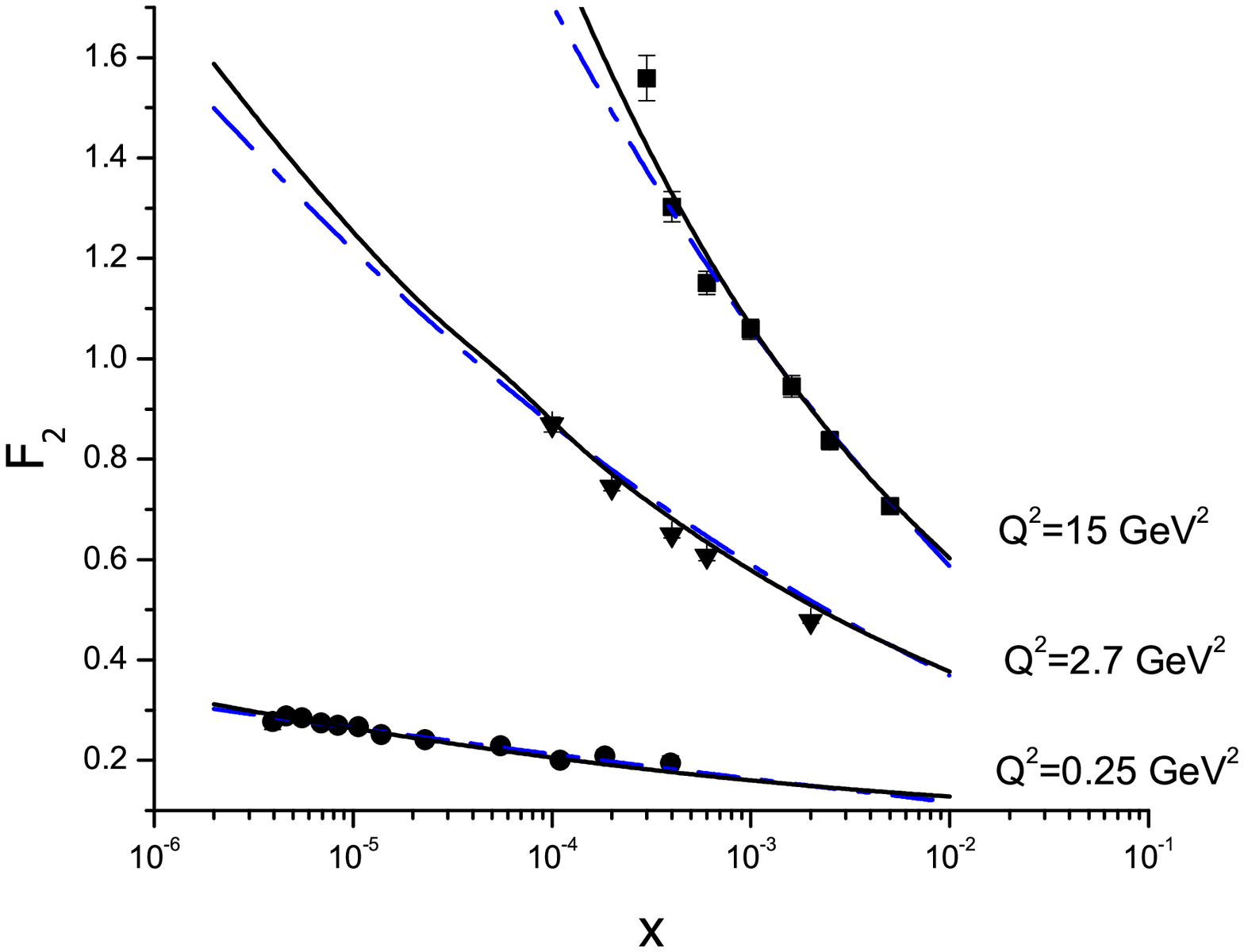}
\caption{Comparison of our new dipole model fits to a subset of DIS data \cite{newf2data}.
Left: No saturation fits. FS2004 Regge dipole fit (dashed line) and
(solid line) a fit 
of the same model to data in the restricted range $5 \times 10^{-4} < x <
10^{-2}$, extrapolated over the whole $x$-range $x < 0.01$.  
Right: Saturation fits. FS2004 saturation fit (solid line) and the CGC dipole
model (dot-dashed line)} 
\label{f2fits}
\end{center}
\end{figure}

The best fit obtained with what we shall henceforth call our FS2004 Regge fit 
is shown as the dashed line in Figure \ref{f2fits} (left) and the parameter
values are listed in 
Table \ref{tab:fs2004regge}. While the values of
the Regge  exponents $ \lambda_S$ and  $\lambda_H$, and of the boundary parameters
$r_0$ and $r_1$ are eminently sensible, the quality of the fit is not
good, corresponding to a $\chi^2$/data point  of 428/156. This is not just a failing of
this particular parameterisation. We have attempted to fit the data with  other Regge
inspired models, including the original FKS parameterization, without success.

A possible reason for this failing is suggested by Figure
\ref{f2fits} (left). At fixed-$Q^2$, the poor $\chi^2$ arises because the fit has much too flat
an energy  dependence at the larger $x$-values for all except the lowest $Q^2$
value. This  could be corrected by increasing the
proportion of the hard term, but this necessarily would lead to a  steeper 
dependence at the lower $x$-values at all $Q^2$. This interpretation is confirmed
 by the solid curve in Figure \ref{f2fits} (left), 
which shows a
the result of fitting only to data in the $x$-range $5 \times 10^{-4} < x <
10^{-2} $, and then extrapolating the fit to lower $x$-values, corresponding to
higher energies at fixed-$Q^2$. As can be seen, this leads to a much steeper
dependence at these lower $x$-values than is allowed by the data at all $Q^2$. 
An obvious way to solve this problem is by introducing saturation at high
energies, to dampen this rise.

\begin{table}[htbp]
\begin{center}
%{\bf Table \ref{tab:fs2004param}}
\be
\begin{array}{c|c|c|c} 
\hline
A_H  & 0.650 & \lambda_H & 0.338  \\
A_S  & 58.42  & \lambda_S & 0.0664 \\
r_0  & 0.872 & r_1       & 4.844   \\
\hline
m    & 0.223 &          &   \\
\hline  
    \end{array}
\ee
\caption{Parameters for the FS2004 Regge model 
in the appropriate GeV based units.}
\label{tab:fs2004regge}
\end{center}
\end{table}

We note that the aforementioned problem could
also be resolved by restricting the fitted region to $x \le 10^{-3}$. In our
view, there is no justification for this, since the non-diffractive contributions
are already small at $x \approx 0.01$. Furthemore,
they make a positive contribution which decreases in size  as $x$-decreases so
that, to the extent that they are not completely negligible, they make the 
problem worse\footnote{For a simple paramaterization of these non-diffractive
contributions, see Donnachie and Landshoff \cite{DL94}.}.

\subsection{Saturation fits to DIS.}

Saturation can be introduced into our new model by adopting a device previously utilized
in \cite{MFGS2000}. Instead of taking $r_0$ to be fixed
we now determine it to be the value at which the hard component is some fixed
fraction  of the soft component, i.e. 
\be
\sigma(x_m,r_0)/\sigma(x_m,r_1) = f
\ee 
and treat $f$ instead of $r_0$ as a fitting parameter. This introduces no new
parameters compared to our previous fit. However, the scale $r_0$ now moves to lower
values as $x$ decreases, and the rapid growth of the dipole cross-section at a
fixed value of $r$ begins to be damped as soon as $r_0$ becomes smaller than $r$. In this sense we
model saturation, albeit crudely, with $r_0$ the saturation radius.  

The best fit  parameter values for what we refer to as our FS2004 saturation fit 
are listed in Table \ref{tab:fs2004sat} corresponding to a $\chi^2$/data point of 155/156.
The corresponding fits to the data are shown in Figure \ref{f2fits} (right). Also shown
are the very similar results obtained using the more sophisticated CGC model (no charm
fit) \cite{IIM04}. 

\begin{table}[htbp]
\begin{center}
\be
\begin{array}{c|c|c|c} 
\hline
A_H  & 0.836 & \lambda_H & 0.324  \\
A_S  & 46.24  & \lambda_S & 0.0572 \\
r_1  & 4.48  & f         & 0.129   \\
\hline
m    & 0.140 &           &   \\
\hline  
\end{array}
\ee
\caption{Parameters for the FS2004 saturation model 
in the appropriate GeV based units.}
\label{tab:fs2004sat}
\end{center}
\end{table}

It is clear from these results that
the introduction of saturation into the model immediately removes the tension between
the soft and hard components which is so disfavoured by the data. However, it is important
to note that this conclusion relies on the inclusion of the data in the low $Q^2$ region: 
both the Regge and saturation models yield satisfactory fits if we restrict to 
$ Q^2 \ge 2 \, {\rm GeV}^2$, with  $\chi^2$/data point values of 78/86 and 68/86
respectively.

\subsection{Predictions for DDIS}

We finally seek support for our suggestion that saturation dynamics may already have
been observed in the HERA data by turning to a comparison to data on diffractive
deep inelastic scattering (DDIS).

In principle, if evidence for saturation is seen in DIS data, it should also be
detectable at some level in DDIS data. As we have seen, there is a
characteristic difference in the predictions of the Regge and saturation models
for the energy dependence of the  DIS structure function $F_2$ at fixed
$Q^2$. If DIS and DDIS are described by the same dipole cross-section, then
there should be corresponding differences in the predictions for the energy
dependence of the diffractive structure function $\FD$ at fixed $Q^2$ and fixed
diffractively produced mass $M_X$. In the modern parlance, we shall examine
the $x_P = (Q^2+M_X^2)/(Q^2+s)$ distribution at fixed $\beta = Q^2/(M_X^2 +
Q^2)$ and fixed $Q^2$.  

A recent discussion of DDIS in the context of the dipole model, including the
predictions of the CGC model, has been presented in reference \cite{FSS04b} and we
refer to that paper for the relevant formulae and more detailed discussion.

Our predictions involve no adjustment of the dipole cross-sections used to
describe the $F_2$ data. However, we are free to adjust the forward slope for
inclusive diffraction, $b$, within the range acceptable to 
experiment and in all cases we took $b=4.5$~GeV$^{-2}$. This simply influences the 
overall normalization of $\FD$ which is therefore free to vary slightly. We are also
somewhat free to vary the value of $\alpha_s$ used to define the normalization
of the $q\bar{q}g$ component, which is important at low values of $\beta$. In all
cases we choose $\alpha_s = 0.1$. 

\begin{figure}[htbp]
\begin{center}
\includegraphics*[width=11cm]{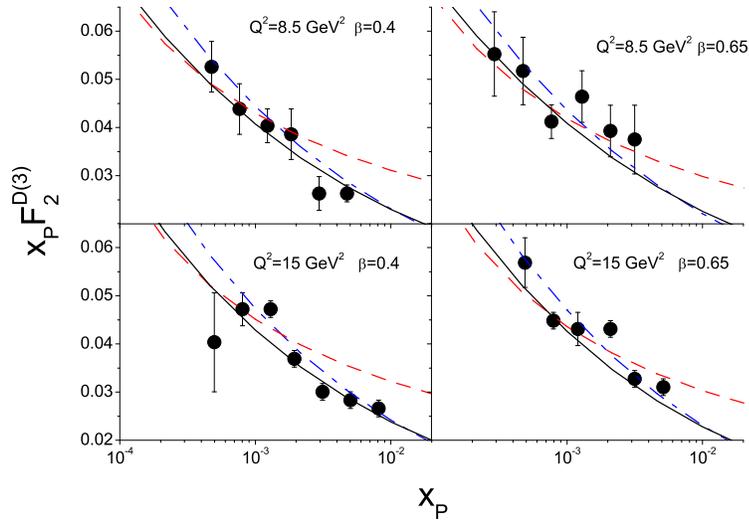}
\caption{
Comparison of the predictions of the FS2004 Regge fit (dashed line), the
FS2004 saturation fit (solid line) and CGC fit (dot-dashed line) to the data on
$F_2^{D(3)}$. Preliminary data from \cite{H103}.
}
\label{fig:F2D3xpom}
\end{center}
\end{figure}

\begin{figure}[htbp]
\begin{center}
\includegraphics*[width=9cm]{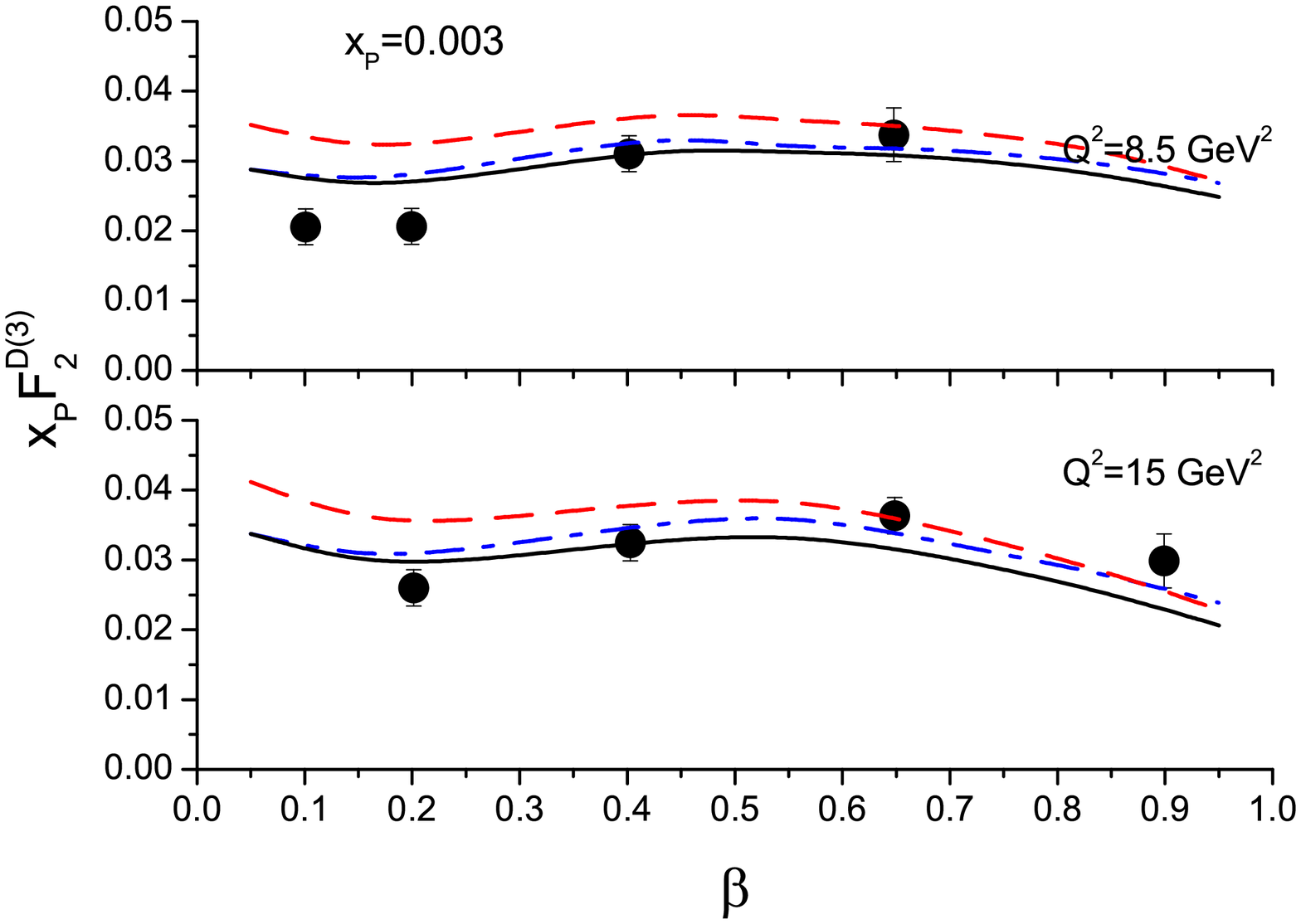}
\caption{
Comparison of the predictions of the FS2004 Regge fit (dashed line), the
FS2004 saturation fit (solid line) and CGC fit (dot-dashed line) to the data on
$F_2^{D(3)}$. Preliminary data from \cite{H103}.
}
\label{fig:F2D3beta}
\end{center}
\end{figure}

In Figure \ref{fig:F2D3xpom}  we show the
predictions of our new FS2004 Regge and saturation models 
for the $x_P$ dependence of the structure function $\FD$ at fixed $Q^2$ and
$\beta$, together with the corresponding predictions of the CGC model.
In doing so, we have chosen to focus on $\beta$
values in the intermediate range where the predictions are relatively
insensitive to the   q\={q}g term and to the  
large $r$ behaviour of the photon wavefunction, which are both rather uncertain. 

There is, as expected, a 
characteristically different energy dependence of the Regge model and the two
saturation models. There is a hint that the data prefer the saturation models, but
more accuracy would be needed in order to make a more positive statement.

In Figure \ref{fig:F2D3beta} we show the $\beta$ dependence at fixed $x_P$ and
$Q^2$. Although this is unlikely to exhibit saturation effects in a transparent
way, it is nonetheless a significant test of the dipole models discussed,  because  
different size dipoles  enter in markedly different  relative weightings to the DIS case.
For $\beta \ge 0.4$, where the $q\bar{q}$ dominates, it is clear that both the Regge
dipole and both saturation models are compatible with the data, given the
uncertainty in normalisation associated with the value of the slope parameter
alluded to above. At the lowest $\beta$ values, where the model dependent
$q\bar{q} g$ component is dominant, there is a discrepancy between the data and
the predictions of all three models, which can only be removed by reducing 
the  value of $\alpha_s$ well below the value assumed \footnote{The relative
size of the $q\bar{q}$ and $q\bar{q} g$ components is given in more detail in
Figure 6 of \cite{FSS04b},}.

\section{Conclusions}

The colour dipole model offers a unified picture of diffractive photoprocesses
over a wide range of $Q^2$. In this paper, we have attempted to throw light on
the question of saturation in dipole models using both DIS and DDIS data.

To do this, we have used a particular dipole cross-section in which saturation 
effects can be included, or not included, in a simple way. When we speak
of saturation we refer specifically to the effect whereby the rapid increase in 
energy of the contribution associated with small dipoles is damped at 
high enough energies. If there is no such damping, we say there is no saturation.
Without saturation, we find that the model is unable to give a satisfactory
account of the data, although the fit is better than with our original Regge
dipole model \cite{FKS99} 
and many variations on it. Furthermore the discrepancies between data and fit
are qualitatively of the sort which one might expect that saturation effects
could remove. On incorporating such effects, we indeed obtain a very good fit to
the data, without extra parameters, and the fit 
is of comparable quality to that obtained using the Colour Glass Condensate 
Model \cite{IIM04}, which incorporates a more
sophisticated treatment of saturation effects. Both models yield
very similar, successful, predictions for the DDIS data in the regions where
the $q \bar{q}$ dipole contribution dominates. 

Finally, we note that the evidence we have presented for gluon saturation rests upon the
applicability of the dipole model. It is of course well known that DGLAP fits to
the DIS data work well down to $Q^2 \sim 1$ GeV$^2$ \cite{MRST,CTEQ}.
Such observations are in accord with our observation that the
higher $Q^2$ data can be fitted without invoking saturation, and it is only on
admitting the lower $Q^2$ data that the evidence emerges. However, Thorne has recently
pointed out that DGLAP corrections to the dipole formulation are surely needed in order
to make precise comparison to the data and therefore that any
claims for saturation are to be viewed with caution \cite{Thorne}. 
This may well be a valid
point and it will be interesting to see if our statement that saturation is
responsible for alleviating the tension between the low and intermediate $Q^2$ data
survives a more careful analysis.

\section{Acknowledgement}
This research was supported in part by a UK Particle Physics and Astronomy 
Research Council grant number PPA/G/0/2002/00471. We would like to thank  Ruben
Sandapen fruitful discussions. 
.

\end{document}